\documentclass[%
reprint,
groupedaddress,
 amsmath,amssymb,
 aps,
pra,
]{revtex4-2}

\usepackage{placeins}
\usepackage{graphicx}
\usepackage{dcolumn}
\usepackage{bm}

\begin{document}

\preprint{APS/123-QED}

\title{Optimal dynamical stabilization}

\author{Arnaud Lazarus}
\email{arnaud.lazarus@upmc.fr}
\affiliation{%
Institut Jean Le Rond d'Alembert, CNRS UMR7190, Sorbonne Universit\'e Paris, France
}
\affiliation{
Massachusetts Institute of Technology, Department of Mathematics, Cambridge, MA 02139, USA
}

\author{Emmanuel Tr\'elat}
\affiliation{
Sorbonne Universit\'e, CNRS, Universit\'e Paris Cit\'e, Inria, Laboratoire Jacques-Louis Lions (LJLL), F-75005 Paris, France
}%

\date{\today}

\begin{abstract}
Minimal conditions for dynamic stability refer to the necessary criteria that system parameters must meet to ensure its response to disturbances remains bounded over time. Lyapunov's and Kapitza's stability criteria are classical examples that have led to major technological advances and fundamental insights in physics. Here, we establish new minimal stability conditions for systems whose potential energy curvature varies periodically in time, extending Kapitza stabilization to what we term optimal dynamical stabilization. Using optimal control theory, we determine the minimal time-periodic stiffness required to stabilize a linear mass-spring system and validate our predictions with model experiments. When the potential curvature alternates between negative and positive values, the set of modulation functions ensuring optimal stability becomes discrete and is governed by eigenvalue problems mathematically analogous to those describing quantum bound states in one-dimensional potential wells. These findings deepen our understanding of dynamical systems and lay the foundation for promising new passive stabilization techniques in applied physics.
\end{abstract}

\maketitle


\section{\label{sec:intro}Introduction}

Kapitza stabilization is a fundamental passive technique in physics that consists of periodically varying a system's properties in time to dynamically sustain one of its naturally unstable configurational states \cite{stephenson1908xx,kapitsa1951pendulum,landau1960mechanics}. By modulating gravitational acceleration, pendulums can be stabilized in their upside-down positions \cite{bogdanoff1965experiments,smith1992experimental,acheson1993upside,richards2018microscopic}, and buoyancy can be inverted so that toy boats eventually float at the bottom surface of a levitating liquid \cite{apffel2020floating}. Periodically varying the electromagnetic or electric field induced by a Josephson junction or a quadrupole coil, respectively, enables the reorientation of a nano-magnet's easy axis \cite{kulikov2022kapitza} or the trapping of charged particles \cite{paul1990electromagnetic}, a key component of mass spectrometers \cite{marshall2008high} and trapped ion quantum computers \cite{bruzewicz2019trapped}. Kapitza stabilization also lies at the core of Floquet engineering \cite{bukov2015universal,yin2022floquet}, which aims to dynamically control exotic quantum states in materials by rapidly modulating their electronic properties using time-periodic external fields such as intensive lasers or electric fields \cite{wei2018pendular}.  
\\
\indent
Over the years, researchers have established the necessary conditions for dynamically stabilizing a diverging mass in the Kapitza limit \cite{kapitsa1951pendulum,landau1960mechanics,bukov2015universal} but the minimal amount of local positive curvature of the potential energy required for stabilization over time-reffered to as optimal dynamical stabilization-remains a broader problem that has so far only been partially explored \cite{lazarus2019discrete,grandi2023new}.
We propose to tackle this stability problem through optimal control theory \cite{kirk2004optimal,trelat2012optimal} combined with model experiments. 
In this article, we focus on a simple mass-spring linear oscillator with a $T$-periodic time-varying stiffness and ask: what is the minimal stiffness required to stabilize the mass? To validate this theoretical framework, we design a model experiment that examines the minimal time-periodic magnetic variation a compass needs for its magnetic north pole to continuously point towards the Earth's magnetic north pole . 
\\
\indent
We find that the optimal control for this problem is bang-bang, and that its solutions can be derived analytically. In the large $T$ limit in particular, the optimal control problem becomes mathematically equivalent to calculating the ground state of a quantum particle confined in a finite square-well potential, with time taking the place of space. More generally, when searching for all bounded oscillations of the mass-spring system under minimal stiffness constraints, the admissible controls for stabilization become quantized and can be predicted by the eigenstates of a stationary Schr\"odinger-like equation. Although the parameter space is found numerically to be remarkably narrow, optimal dynamical stabilization is successfully achieved in our experiments, with measurements closely matching theoretical predictions.
This article provides fresh insights into the stability of dynamical systems and the mathematics of linear ordinary differential equations. Since the concept is independent of the scale or nature of the modulation, it paves the way for improved passive control techniques in physics by fully leveraging the mathematical analogy with the one-dimensional  Schr\"odinger equation.

\section{Theory}

\begin{figure}[!t]
\centering
\includegraphics[width=0.98\columnwidth]{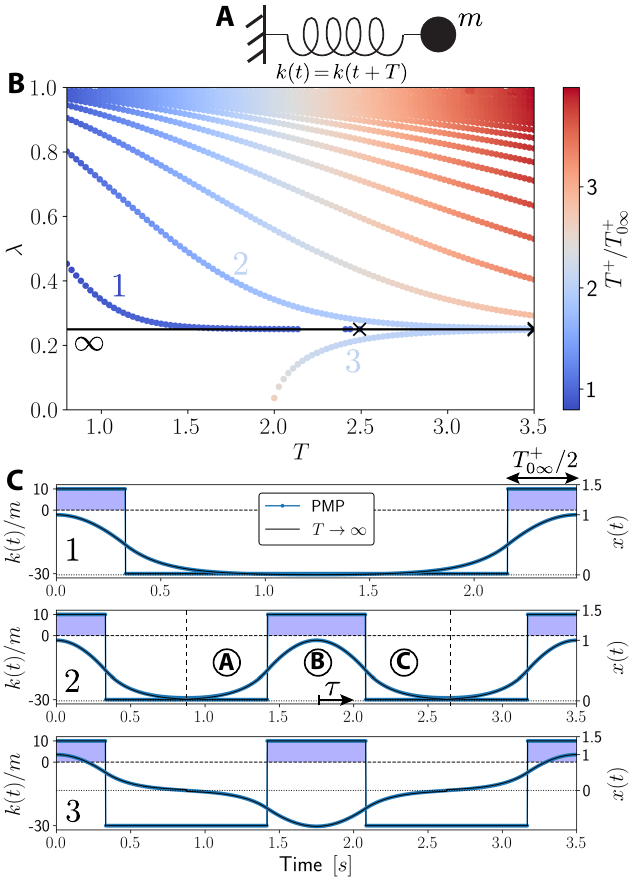}
\caption{Optimal "bang-bang" control of the $T$-periodic solutions of a mass in a time-varying harmonic potential. A. Linear mass-spring oscillator model with a periodically varying stiffness $k(t)=k(t+T)$ with $10 \leq K/m \leq 30$ s$^{-2}$. B. Solutions of the PMP problem Eqs.(\ref{Pontry1reduced})-(\ref{Pontry2reduced}) in the $(\lambda,T)$ space for $x_0=1$ and $y_0=0$. The color bar shows the ratio between the duration $T^+$ of $u(t)=u^+$ and the duration $T^+_{0\infty}$ given in Eq.(\ref{T+sym}) in the large $T$ limit. The thin black line represents the analytical solution $\lambda_{\infty}=\cos^2(\sqrt{u^+}T^+_{0\infty}/2)$. C. Comparison between typical optimal numerical responses $x(t)$ in blue line and analytical responses $x_{0\infty}(t)$. The optimal control $u(t)$ is also shown. Solution $x_1(t)$, shown for $T=2.5$ s, is the globally optimal solution, when $x_2(t)$ and $x_3(t)$, plotted for $T=3.5$, are locally optimal solutions.}
\label{Figure0_article_opti_control}
\end{figure}

The fundamental model to get physical insights in dynamical stabilization is the linear mass-spring system of Fig.\ref{Figure0_article_opti_control}A where $m$ is the mass and $x(t)$ its unidimensional motion as a function of time. The only difference with a classic harmonic oscillator is that the spring stiffness $k(t)$ is $T$-periodic in time. For simplicity but without lack of generality, we assume the stiffness is contained between $K>0$ and $K-\Delta K<0$.  The goal of optimal dynamical stabilization is to find what is the minimal amount of stiffness for a non-trivial response $x(t) \neq 0$ to still be bounded as time goes to infinity (we recall that whatever $T$, if $k(t)<0 \; \forall t$, $x(t) \rightarrow +\infty$ when $t \rightarrow +\infty$ if $x(0) \neq 0$). 

The motion of the mass is simply governed by the following second order linear ordinary differential equation (ODE)
\begin{equation}
\label{eqmov}
\ddot{x}(t)+u(t)x(t)=0\end{equation}
subjected to the initial conditions $x(0)=x_0$ and $\dot{x}(0)=y_0$ and where $u(t)=u(t+T)=k(t)/m$ is a $T$-periodic square wave function contained between $u^+ = K/m > 0$ and $u^- = K/m-\Delta K/m <0$. With these notations, optimal dynamical stabilization consists in finding what is the minimal $\int_0^Tu(t)dt$ for which a non trivial $x(t)$ is bounded. To start with this problem, we focus on modulation function $u(t)$ associated with $T$-periodic solutions $x(t)=x(t+T)$ that we know from Floquet theory \cite{whittaker1996course,richards2012analysis, lazarus2015stability} are primordial in structuring the linear time-periodic system given in Eq.(\ref{eqmov}).

\subsection{Optimal control of $T$-periodic solutions}

The optimal problem is now to minimize the continuous-time functional $\int_0^Tu(t)dt$ with $u^- \leq u(t) \leq u^+$ subjected to the linear ODE in eq.(\ref{eqmov}) and the endpoints conditions $x(0)=x(T)$ and $\dot{x}(0)=\dot{x}(T)$. Applying Pontryagin's maximum principle (PMP), one can actually prove (see the full demonstration in \cite{Trelat2025} and a summary of it in the SI) there is a unique non trivial optimal trajectory to this problem such that 
\begin{equation}
\left\{\begin{array}{ll}
\dot{x}(t)= y(t) \\
\dot{y}(t)= -u(t)x(t)\end{array}\right.
\label{Pontry1reduced}
\end{equation}
with $x(0)=x(T)=1$, $y(0)=y(T)=0$ and 
\begin{equation}
u(t)=\left\{\begin{array}{ll}
u^-<0 \quad \text{if } x^2(t) < \lambda\\
u^+ > 0 \quad \text{if } x^2(t) > \lambda \end{array}\right.
\label{Pontry2reduced}
\end{equation}
where the ``shooting parameter'' $\lambda$ is a real number and the only unknown of the PMP. There is actually an infinite number of optimal solutions verifying $\mu (x(0)^2 + y(0)^2) = 1$ with $\mu \neq 0$ because of homothety and phase independence of the problem but all are associated with the same value of $\int_0^Tu(t)dt$ and we can assume $x(0)=1$ and $y(0)=0$ with no loss of generality. Furthermore, the globally optimal control $u(t)$ is bang-bang with two switchings and
\begin{equation}
u(t)=\left\{\begin{array}{lll}
u^+ \quad \text{if }  \, 0 \leq t < T^+/2\\
u^- \quad \text{if } \, T^+/2 < t < T-T^+/2\\
u^+ \quad \text{if } \, T-T^+/2 < t < T \end{array}\right.
\label{Control}
\end{equation}

The PMP problem given in Eqs.(\ref{Pontry1reduced})-(\ref{Pontry2reduced}) can be solved for a given $T$, by a shooting method over $[0,T]$. Numerical results are shown in Fig.\ref{Figure0_article_opti_control}B-C where we ran $60000$ shooting methods between $T=0$ and $3.5$ s and some initial guesses of $\lambda$ between $0.01$ and $0.99$. The branch of $\lambda$ associated with the globally optimal solution of Eqs.(\ref{Pontry1reduced})-(\ref{Pontry2reduced}) is denoted by $1$ in Fig.\ref{Figure0_article_opti_control}B; the corresponding solution $x(t)$ and its bang-bang control verifying Eq.(\ref{Control}) are given in Fig.\ref{Figure0_article_opti_control}C1. The optimal solutions exhibit a turnpike property \cite{trelat2012optimal}: $x(t)$ approaches $ 0^+$ with $u(t)=u^-$ for the major part of the period $T-T^+$ to minimize $\int_0^Tu(t)dt$ until it eventually departs to fulfill the non trivial periodicity condition $x(t)=x(t+T)$ during the time $T^+$ where $u(t)=u^+$. Interestingly in the large $T$ limit, $\lambda$, the duration $T^+$ of $u(t)=u^+$ and the globally optimal $x(t)$ converge towards $\lambda_{\infty}$, $T^+_{0\infty}$ and $x_{0\infty}(t)$, respectively, that can be analytically determined as explained in the SI: $\lambda_{\infty}=\cos^2(\sqrt{u^+}T^+_{0\infty}/2)$, $T^+_{0\infty}$ is the smallest $T^+_{\infty}$ of the transcendental equation
 \begin{equation}
\sqrt{|u^-|}=\sqrt{u^+}\tan(\sqrt{u^+}T^+_{\infty}/2),
\label{T+sym}
\end{equation}
and 
\begin{align}
\left\{\begin{array}{lll}
x_{0\infty}(\tau) = \frac{\cos(\sqrt{u^+}T^+_{0\infty}/2)}{e^{-\sqrt{|u^-|}T^+_{0\infty}/2}}e^{\sqrt{|u^-|} \tau} & \text{for } \tau \in A\\
x_{0\infty}(\tau) = \cos(\sqrt{u^+} \tau)  & \text{for } \tau \in B\\
x_{0\infty}(\tau) = \frac{\cos(\sqrt{u^+}T^+_{0\infty}/2)}{e^{-\sqrt{|u^-|}T^+_{0\infty}/2}}e^{-\sqrt{|u^-|} \tau} & \text{for } \tau \in C\end{array}\right.
\label{boundstatemanu}
\end{align}
where the local time $\tau$ starts at the middle of each $T^+$ region and is partitioned in three such that: $A = \{ \tau \mid \tau \in [-T/2, -T^+/2] \}$, $B = \{ \tau \mid \tau \in [-T^+/2, T^+/2] \}$ and $C = \{ \tau \mid \tau \in [T^+/2,T/2] \}$ as shown in Fig.\ref{Figure0_article_opti_control}C. Those analytical quantities show excellent agreement with the globally optimal solution of the PMP already from $T=2.5$ s as shown in Fig.\ref{Figure0_article_opti_control}B-C1. Note that Eqs.(\ref{T+sym})-(\ref{boundstatemanu}) are well known by physicists because they represent the ground state of a quantum particle confined in a finite potential well, albeit with space replaced by time \cite{messiah1961quantum}.

An infinite number of solutions of Eqs.(\ref{Pontry1reduced})-(\ref{Pontry2reduced}) exists on top of the globally optimal one as shown by the many branches of solutions in Fig.\ref{Figure0_article_opti_control}B. Those locally optimal solutions can be qualitatively sorted based on the number of switchings of the control $u(t)$ and the symmetries of $x(t)$ over $[0,T]$ \cite{Trelat2025}. Fig.\ref{Figure0_article_opti_control}C2 ($4$ switchings) and C3 ($4$ switchings but $x(t)$ can be negative) shows the solutions $x(t)$ of the branches $2$ and $3$, respectively, for $T=3.5$ s. In the large $T$ limit, the shooting parameter $\lambda$, the durations $T^+$ and the locally optimal solutions $x(t)$ can eventually be derived from the aforementioned $\lambda_{\infty}$, $T^+_{0\infty}$ and $x_{0\infty}(t)$ given in Eqs.(\ref{Control})-(\ref{boundstatemanu}) as shown by the thin black lines of Figs.\ref{Figure0_article_opti_control}B-C.

\begin{figure*}[!t]
\centering
\includegraphics[width=.9\textwidth]{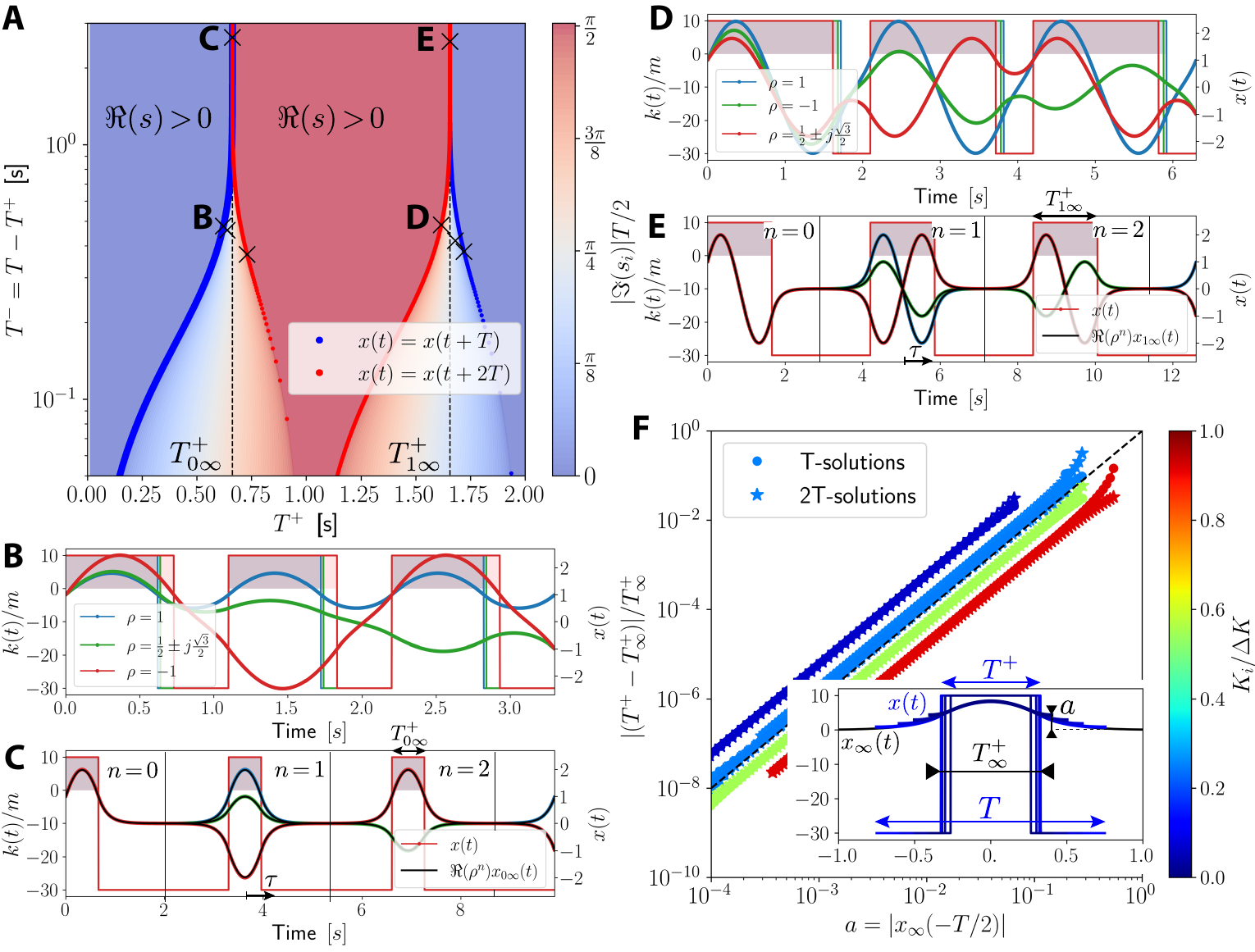}
\caption{Local stability analysis of $\ddot{x}(t)+u(t)x(t)=0$ with $u(t)= u^+=10$ s$^{-2}$ during $T^+$ and $u(t)=u^-=-30$ s$^{-2}$ the rest of the period $T^-=T-T^+$. A. Evolution of the imaginary part of the Floquet exponent in the $(T^+,T^-)$ space. B-C. Modulation function $u(t)$ and their associated responses $x(t)$ in the first ``optimal'' stability region about $T^+ \approx T^+_{0\infty}$ for three stable trajectories corresponding to a $T$-periodic ($\rho=1$), $2T$-periodic ($\rho=-1$) and a quasi-periodic ($\rho = 1/2 + j\sqrt{3}/2$) case. Two different periods are shown: $T=1.1$ s in B and $3.3$ s in C. D-E. Same as B-C. but in the second stability region about $T^+ \approx T^+_{1\infty}$. The period in D is $T=2.1$ s and $T=4.2$ s in E. F. Convergence of the duration $T^+$ of $u(t)=u^+$ associated with $T$ and $2T$-periodic solutions towards their $T^+_{\infty}$ counterparts in the large $T$ limit for various $K$ and $\Delta K$. Inset describes how the $T$-periodic $x(t)$ and $u(t)$ converge towards $x_{\infty}(t)$ and $T^+_{\infty}$ in the first ``optimal'' stability region and explains the meaning of the quantity $a=x_{\infty}(T/2)$.}
\label{fig1}
\end{figure*}

\subsection{Full stability problem}

We now proceed by studying the full region of bounded solutions $x(t)$ of the original equation of motion Eq.(\ref{eqmov}), $\ddot{x}(t)+u(t)x(t)=0$. This time, we impose the $T$-periodic modulation function $u(t)$ and compute the response $x(t)$. As suggested by the previous section, we focus on $u(t)$ with two switchings that read $u(t)=u^+>0$ during $T^+$ and $u(t)=u^-<0$ during the rest of the period $T^-=T-T^+$. According to Floquet theory, we know that, for any $u(t)$, the solution $x(t)$ takes the normal form
\begin{equation}
\label{periodicmapping}
x(t) = \Psi(t)e^{st} + \bar{\Psi}(t)e^{-st}\end{equation}
where $\Psi(t)=\Psi(t+T)$ is a complex function with the same periodicity as $u(t)$ ($\bar{\Psi}(t)$ is the complex conjugate of $\Psi(t)$) and $s$ is a complex number called the Floquet exponent. For a given $u(t)$, the stability of the mass is usually assessed by looking at the Floquet exponent or the Floquet multiplier $\rho=e^{sT}$. Fig.\ref{fig1}A is typical of linear Floquet systems and shows the evolution of the imaginary part of the Floquet exponent $s$ in the $(T^+,T^-)$ modulation space for $u^+=10$ s$^{-2}$ and $u^-=-30$ s$^{-2}$. As expected from 1 degree-of-freedom linear time-periodic systems, several regions of unbounded solutions $x(t)$ with $\Re(s)>0$ ($|\rho| > 1$) exist that are separated by regions of $\Re(s)=0$ ($|\rho| = 1$) where $x(t)$ remains bounded for all $t$. Those ``stability'' regions with $\Re(s)=0$ are enclosed by $T$-periodic solutions $x(t)=x(t+T)$ with $s=0$, or $\rho =1$ (blue dots in Fig.\ref{fig1}A), and $2T$-periodic solutions $x(t)=x(t+2T)$ with $s=i\pi/T$, or $\rho=-1$ (red dots in Fig.\ref{fig1}A). Between  $s=0$ and $s=i\pi/T$, an infinite number of bounded quasi-periodic solutions are found. As the duration $T^-$ of $u(t)=u^-$ increases, minimizing the quantity $\int_0^Tu(t)dt$, the width of the ``stability'' regions decreases about a discrete set of durations $T^+_{i \infty}$ with $i \in \mathbb{Z}^+$ and $T^+_{(i+1)\infty} > T^+_{i \infty}$. We first focus on the first ``optimal'' region associated with the smallest $T^+_{i \infty}$, denoted $T^+_{0 \infty}$, that we already derived in Eq.(\ref{T+sym}).

Figs.\ref{fig1}B-C show the bounded trajectories $x(t)$ with $|\rho|=1$ and their associated modulation functions $u(t)$ as a function of time for $\rho=1$ ($T$-periodic solution), $\rho=-1$ ($2T$-periodic solution) and $\rho=1/2+j\sqrt{3}/2$ (quasi-periodic response) in the optimal stability region $T^+ \approx T^+_{0\infty}$ for $T=1.1$ and $T=3.3$ s. As $T^-=T-T^+$ increases for example from Fig.\ref{fig1}B to \ref{fig1}C, we see that all the $u(t)$ associated with bounded solutions converge towards a unique $u(t)$ with $T^+ \rightarrow T^+_{0\infty}$. Moreover, as shown in thin black lines in Fig.\ref{fig1}C, all those bounded trajectories with Floquet multiplier $|\rho|=1$ can now be expressed as $x_n(t)=x_{0\infty}(t)\Re(\rho^n)$ where $x_n(t)$ is the trajectory over the $n^{th}$ period with $n \in \mathbb{N}_0$ and $x_{0\infty}(t)$ is the $T$-periodic optimal solution in the large $T$-limit already expressed in Eq.(\ref{boundstatemanu}). In other words, as $T$ increases, the sole knowledge of $T^+_{0\infty}$ and $x_{0\infty}(t)$ becomes sufficient to fully characterize all the optimal bounded solutions of $\ddot{x}(t)+u(t)x(t)=0$ characterized by a Floquet multiplier $|\rho|=1$. As shown in the inset of Fig.\ref{fig1}F, one could get an estimate of when the reduction of the full stability problem to Eqs.(\ref{T+sym})-(\ref{boundstatemanu}) is reasonable by simply imposing the normalization condition $x_{0\infty}(t)=\int_0^Tx_{0\infty}^2(t)dt=1$ and evaluating the quantity $a=x_{0\infty}(T/2)$. As $T$ increases such that $a$ decreases and $x(t)$ becomes compact on each period, $T^+$ and $x(t)$ converge to $T^+_{0\infty}$ and $x_{0\infty}(t)$ in a quadratic fashion with a prefactor close to one, whatever $u^+$ or $u^-$.

\begin{figure}[!h]
\centering
\includegraphics[width=0.85\columnwidth]{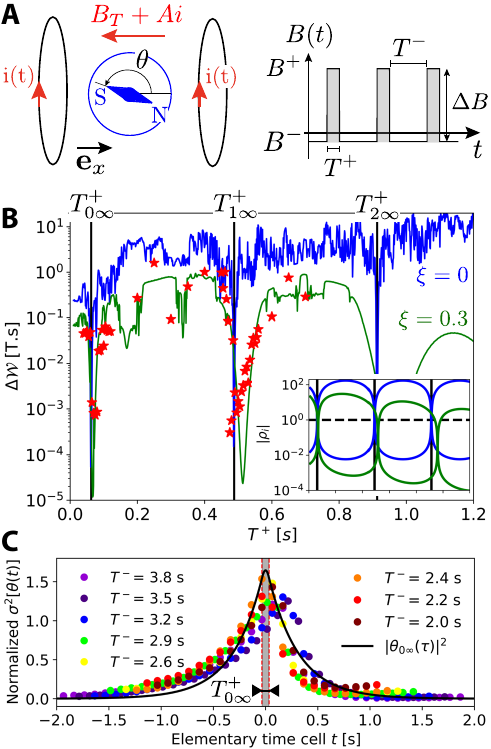}
\caption{Experimental proof of concept. A. A compass, fully parameterized by the angle $\theta(t)$ between $\mathbf{e}_x$ and its $N-S$ local axis is placed at the center of two Helmholtz coils oriented along $\mathbf{e}_x$. We impose a $T$-periodic magnetic field $B(t)=B(t+T)$ with $B(t)=B^-=-B_T=-47$ $\mu$T ($i=0$ A) during a time $T^-$ and $B(t)=B^+=852$ $\mu$T during $T^+$ ($i=-200$ mA) so that $T=T^-+T^+$. B. Evolution of $\Delta \mathcal{W} = \int_0^{6T}|\theta(t) B(t)sin(\theta(t))|dt$ as a function of $T^+$ for $T^-=2.5$ s. Red crosses show experimental data when blue and green lines represent numerical simulations with $\xi=0$ and $\xi=0.3$, respectively. Vertical black lines represent $T^+=T^+_{i\infty}$ computed from Eq.(\ref{eqmovBVP}). Inset shows the evolution of the absolute value of the Floquet multipliers of the numerical stability problem about $\theta=\pi$ as a function of $T^+$.  C. Evolution of the measured normalized variance ($\int_{-T/2}^{T/2}\sigma^2[\theta(t)]=1$) of the oscillation angles about $\theta=\pi$ at various phase of the periodic imposed field $B(t)$ for $T^+=70$ ms (vertical red dashed lines) and various values of $T^-$. The first eigenfunction $\theta^2_{0\infty}(\tau)$ of Eq.(\ref{boundstatemanu}) for $\Delta B = B^+-B^-= 899$ $\mu$T, $T^+_{0\infty}=62$ (shown in grey shade) ms and an eigenvalue $B^+=852$ $\mu$T is shown in black line. }
\label{fig3}
\end{figure}

Another important feature of the full stability problem highlighted in Fig.\ref{fig1}A is that the optimal modulation functions $u(t)$ associated with $T^+ \rightarrow T^+_{0\infty}$ are not the only functions leading to bounded solutions $x(t)$ with small $\int_0^Tu(t)dt$ in the large $T$ limit. When $T \rightarrow \infty$, there is actually a countable infinite set of durations $T^+_{i\infty}$ with $i \geq 1$ that correspond to higher modes of stabilization \cite{grandi2023new}. For example, Figs.\ref{fig1}D-E show the mode of stabilization in the region $T^+ \approx T^+_{1\infty} = 1.65$ s where $x(t)$ have the ``time'' to do two oscillations during $T^+$. Although not optimal in the sense of the PMP problem given in Eqs.(\ref{Pontry1reduced})-(\ref{Pontry2reduced}), these solutions of $\ddot{x}(t)+u(t)x(t)=0$ with $|\rho|=1$ converge towards a unique modulation function $u(t)$ with $T^+ \rightarrow T^+_{i\infty}$ and a bounded trajectory $x_n(t)=x_{i\infty}(t)\Re(\rho^n)$ in the large $T$ limit as shown in Fig.\ref{fig1}E. For $T \rightarrow \infty$, $\ddot{x}(t)+u(t)x(t)=0$ becomes,
\begin{equation}
\label{eqmovBVP}
\left\{\begin{array}{ll}
\left(-\frac{d^2}{d\tau^2}+ \Delta u\right)x_{i\infty}(\tau)=u^+x_{i\infty}(\tau) \text{ if }  |\tau| > T^+_{i\infty}/2\\
\left(-\frac{d^2}{d\tau^2}+ 0\right)x_{i\infty}(\tau)=u^+x_{i\infty}(\tau) \text{ if }  |\tau| < T^+_{i\infty}/2 \\
\end{array}\right.
\end{equation}
with $\Delta u = u^+-u^-$ and where the local time $\tau$ starts at the middle of each $T^+_{i\infty}$ region as shown in Figs.\ref{fig1}C-E. Upon normalization of $x_{i\infty}(\tau)$ and because of the turnpike property of the solution $x(t)$ in the large $T$ limit, we get $x_{i\infty}(-\infty)=x_{i\infty}(+\infty) \rightarrow 0$. Again, Eq.(\ref{eqmovBVP}) is the mathematical analogue of the stationary Schr\"odinger equation of a particle in a finite potential well. For a given ``height'' $\Delta u$, the $T^+_{i\infty}$ are the widths of the potential well that allow $u^+$ to be an eigenvalue of Eq.(\ref{eqmovBVP}) and the functions $x_{i\infty}(t)$ are the associated eigenfunctions. The smallest duration $T^+_{0\infty}$ associated with the first  eigenfunction $x_{0\infty}(t)$ are the optimal solutions of the PMP problem given in Eqs.(\ref{Pontry1reduced})-(\ref{Pontry2reduced}) in the large $T$ limit. As $T^-=T-T^+$ increases, all the modulation functions $u(t)$ and associated bounded responses $x(t)$ of $\ddot{x}(t)+u(t)x(t)=0$ associated with $|\rho|=1$ eventually converge towards the solutions of Eq.(\ref{eqmovBVP}) in a quadratic fashion as shown in Fig.\ref{fig1}F.

\section{Experimental proof of concept}

The experimental proof of concept consists in a compass centered and aligned between two Helmholtz coils as shown in Figure  \ref{fig3}A. The dipole's configurational state is fully parametrized in time by the angle $\theta(t)$ between the axis of the coils $\mathbf{e}_x$, coincident with the North-South magnetic axis of earth, and the $N-S$ axis of the magnetized needle. This classic 1 degree-of-freedom nonlinear oscillator can be modeled by the following equation of motion 
\begin{equation}
\label{eqexpmov}
\ddot{\theta}(t)+2\xi \sqrt{\frac{|\mathbf{B}(i)\mathbf{e}_x|\mu}{I}}\dot{\theta}(t)+\frac{\mathbf{B}(i)\mathbf{e}_x\mu}{I}\sin(\theta(t))=0\end{equation}
where $\xi$ is the damping ratio and $\omega(i)=\sqrt{|\mathbf{B}(i)\mathbf{e}_x|\mu/I}$ is the natural frequency of the dipole about its stable equilibrium position that can be either $\theta(t)=0$ or $\theta(t)=\pi$ depending on the current $i$ in the coils. $\mathbf{B}(i)=-(B_T + A i)\mathbf{e}_x$ is the uniform magnetic field felt by the dipole with $B_T=47$ $\mu$T the earth magnetic field, $A=4496 $ $\mu$T/A a property related to the Helmholtz coils configuration and $i$ the current in the coils. Measuring the natural frequency and damping ratio of the magnetic dipole when perturbed from its stable equilibrium position for various $i$ (see Figure S1), we find $\xi=0.3$ and the ratio between the magnetic moment $\mu$ and moment of inertia $I$ to be $\mu/I=6.4 \times 10^4$ A.kg$^{-1}$. When $i>-B_T/A \approx -10$ mA, the magnetic field is pointing towards $-\mathbf{e}_x$ and so is the north pole of the compass: $\mathbf{B}(i)\mathbf{e}_x<0$ and therefore $\theta =0$ is stable when $\theta=\pi$ is unstable. For a constant current such that  $i < -10$ mA, the stability of the equilibrium configurational states is reversed: a compass with any initial condition eventually ends up towards $\theta = \pi$ in a damped oscillatory fashion and diverges from $\theta = 0$.

\subsection{Optimal dynamical stabilization}
We now focus on the configurational state $\theta(t)=\pi$ that is naturally unstable when $i(t)=0$, i.e. when no power is consumed by the coils. Optimal dynamical stabilization consists in trapping the south pole of the compass towards the magnetic south pole of earth (stabilize $\theta$ about $\theta(t) = \pi$) with a minimal energy, which is here represented by $\int |i(t)|dt$. To do so, we impose a $T$-periodic magnetic field $B(t) =\mathbf{B}(i)\mathbf{e}_x=B(t+T)$ that reads $B(t)=B^-=-47$ $\mu$T during $T^-$ ($i(t)$=0 in the coils) and $B(t)=B^+=+852$ $\mu$T during $T^+$ ($i(t)=-200$ mA) such that $T=T^++T^-$. An electromagnet keeps the compass in $\theta(t)=\pi$ and $\dot{\theta}(t)=0$ until at $t=0$, the electromagnet is turned off and the compass starts to be subjected only to $B(t)$. 

Figure \ref{fig3}B shows the evolution of $\Delta \mathcal{W} = \int_0^{6T}|\theta(t) B(t)sin(\theta(t))|dt$ as a function of $T^+$ for a fixed value of $T^-=2.5$ s with red stars representing experimental data and green and blue lines displaying numerical data obtained by integrating Eq.(\ref{eqexpmov}) with $\xi=0$ and $\xi=0.3$, respectively (and initial conditions $\dot{\theta}(0)=0$ and $\theta(0)=\pi+5 \times 10^{-3}$ as the only fitting parameters). We see very good agreement between experimental data and numerical ones for $\xi=0.3$ (see e.g. Fig.S2 of the SI to see the evolution of the experimental and numerical trajectories over $6$ periods for $T^+=70$ ms and $T^+ = 200$ ms). For most of the $T^+$, the compass is rapidly having large $\theta(t)$ due to the periodic magnetic field $B(t)$ and the quantity $\Delta \mathcal{W}$, that is proportional to the work done by the dipole in the magnetic field, is large. But there are "absorption rays" of $T^+$ where $\Delta \mathcal{W}$ drops drastically because the needle's south pole is dynamically stabilized about $\theta=\pi$. The inset of \ref{fig3}B shows the evolution of the absolute value of Floquet multipliers $|\rho|$ as a function of $T^+$ obtained from the stability analysis of Eq.(\ref{eqexpmov}) about $\theta=\pi$ for $\xi=0$ and $\xi=0.3$. Unlike the undamped case shown in the first section of this paper, the damping breaks the symmetry of $|\rho|$ with respect to $|\rho|=1$ and as a consequence, the width of the stability regions increases and the latter are slightly shifted towards larger $T^+$ as shown in Figure \ref{fig3}B. The vertical black lines, which coincide with the local minima of $\Delta \mathcal{W}$ for $\xi=0$ are the values of $T^+=T^+_{i\infty}$ one finds by solving the eigenvalue problem Eq.(\ref{eqmovBVP}) where now $u^+=B^+\mu/I$, $u^-=B^-\mu/I$ and the eigenfunctions $x_{i\infty}(\tau)\equiv \theta_{i\infty}(\tau)$ should now represent the oscillations of the dynamically stabilized compass. 

Figure \ref{fig3}C focuses on the optimal stability region about $T^+_{0\infty}$ in Figure \ref{fig3}B. When imposing $\Delta B = (u^+ - u^-)\mu/I = 899$ $\mu$T and to obtain a first eigenvalue $u^+ = B^+\mu/I$ with $B^+=852$ $\mu$T in Eq.(\ref{eqmovBVP}), one needs to ensure $T^+_{0\infty} = 62$ ms as shown in grey shade in Figure \ref{fig3}C. This value is in good agreement with experiments where the smallest value of $\Delta \mathcal{W}$ in the first ``absorption ray'' was found at $T^+=70$ ms as represented with vertical red dashed lines in Figure \ref{fig3}C. In the same pulse-centered elementary periodic cell representation, we investigate the predictive power of the eigenfunction $\theta_{0\infty}(\tau)$ of Eq.(\ref{eqmovBVP}) with regard to the stable experimental oscillations $\theta(\tau)$ about $\theta=\pi$ for $T^+=70$ ms. For practical purposes, because the dynamically stabilized motion eventually undergoes a symmetry breaking on each successive modulation period, sensitivity to initial conditions makes it impossible to reproduce the same experiment twice (see e.g. Fig.S3 of the SI) and it is therefore meaningless to directly predict the experimental evolution $\theta(\tau)$. As shown in Figure \ref{fig3}C for $100$ s of recorded motion though (the theoretical and experimental variances have been normalized so that $\int|\theta_{0\infty}(\tau)|^2=1$ and $\int_{-T/2}^{T/2}\sigma^2[\theta(\tau)]=1$), what is conserved throughout experiments is the variance $\sigma^2$ of the experimental trajectories $\theta(\tau)$ on each modulation period which remarkably scales with $|\theta_{0\infty}(\tau)|^2$, because of the property $x_n(t)=x_{0\infty}(t)\Re(\rho^n)$ in the large $T$-limit (see Figs.\ref{fig1}B-C). It is interesting to note that in the optimal stability region for $T^+=70$ ms, we have been able to turn off the coils $3.8$ s, i.e. more than $98 \%$ of the time. The fact that we cannot go to higher $T^-$ in our experiment (for $T^-$ higher than $3.8$ ms, $\theta=\pi$ destabilizes after few periods) is because not only the width of the stability regions about $T^+_{\infty}$ eventually becomes too small according to linear stability analysis like in Figure \ref{fig1}A but the same is true for the basins of attraction about $\theta=\pi$ \cite{grandi2023new}.

\begin{figure}[!t]
\centering
\includegraphics[width=.92\linewidth]{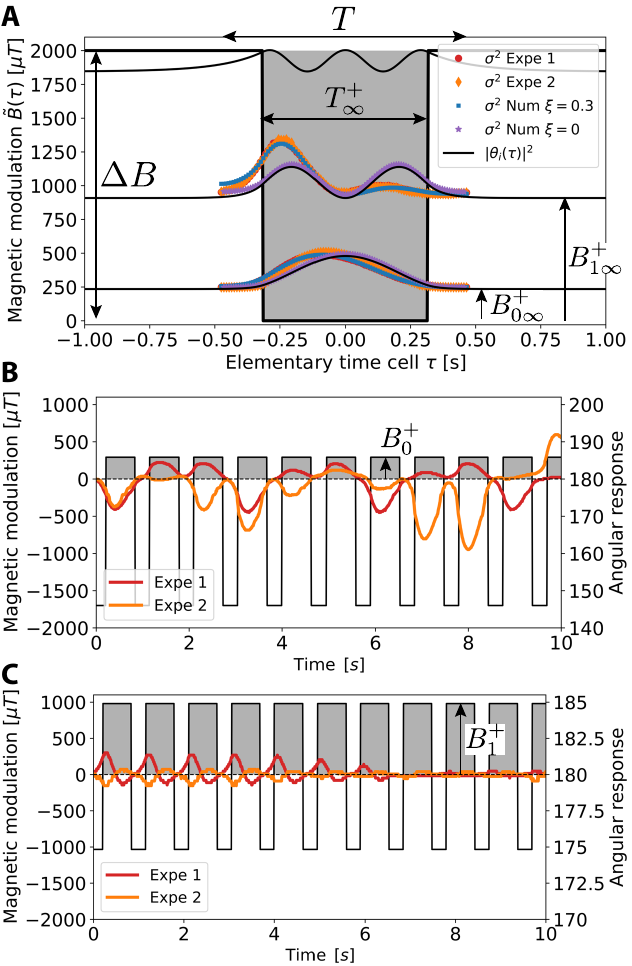}
\caption{Experimental modes of dynamical stabilization. A. Time analog of a quantum particle confined in a finite potential well with $\Delta B = 2000$ $\mu$T, $T^+_{\infty}=630$ ms, $B^+_{0\infty}=235$ $\mu$T, $B^+_{1\infty}=908$ $\mu$T and $B^+_{2\infty}=1845$ $\mu$T. $|\theta_{i\infty}|^2$ with $\theta_{i\infty}$ solution of Eq.(\ref{eqmovBVP}) is shown in black lines with a local origin at $B^+_{i\infty}$. Color dots represent the variance of the experimental and numerical trajectories with $\xi=0$ and $\xi=0.3$ for $T^+=630$ ms, $T^-=320$ ms and i) $B^+=245 \pm 5$ $\mu$T and $B^-=-1746 \pm 5$ $\mu$T for the first mode and ii) $B^+=933 \pm 5$ $\mu$T and $B^-=-1081 \pm 5$ $\mu$T for the second mode. B. Evolution of two stable trajectories with the same experimental initial conditions corresponding to the first mode of optimal stabilization. C. Same as B but for the second mode of stabilization.}
\label{fig4}
\end{figure}

\subsection{Analogue of a particle in a finite potential well}

Following the analogous Schr\"odinger equation in Eq.(\ref{eqmovBVP}), higher modes of stabilization should exist besides the ``ground'' state shown in Fig.\ref{fig3} that corresponds to the minimal $T^+$ one can impose for a given $B^+$ and $\Delta B = B^+-B^-$ as $T$ becomes large. Imposing $\Delta u=u^+ - u^-=\Delta BI/\mu$ with $\Delta B = 2000$ $\mu$T and $T^+_{\infty}=630$ ms, Eq.(\ref{eqmovBVP}) gives three bound states associated with an eigenfunction $\theta_{i\infty}$ and an eigenvalue $B^+_{i\infty}$ such that $B^+_{0\infty}=235$ $\mu$T, $B^+_{1\infty}=908$ $\mu$T and $B^+_{2\infty}=1845$ $\mu$T, as shown in Fig.\ref{fig4}A. 

To confirm that at least two modes of stabilization are possible in our experiment in the large $T$ limit, we fix $T^+=630$ ms and $T^-=320$ ms ($T^-=320$ ms is the maximum duration before the compass destabilizes, hence the so-called experimental large $T$ limit) and impose a $B^+$ and $B^-$ to respectively match the aforementioned theoretical $B^+_{i\infty}$ and $B^-_{i\infty}=B^+_{i\infty}-\Delta B$. We find the compass is stabilized about $\theta=\pi$ on its first mode when imposing $B^+=245 \pm 5$ $\mu$T and $B^-=-1746 \pm 5$ $\mu$T, with a trajectory having one maxima per period as shown in Fig.\ref{fig4}B. Because the dipole is undergoing a symmetry breaking every period in the large $T$ limit, two experiments with the same initial conditions will give two different trajectories, as illustrated by the red and yellow dots in Fig.\ref{fig4}B. However, the variance of those trajectories at various phase along an elementary periodic cell remains identical upon normalization as shown in Fig.\ref{fig4}A where we chose $\int_{-T/2}^{T/2}\sigma^2[\theta(t)]=1$. The experimental variance is well predicted by the variance $\sigma^2[\theta(t)]$ of the numerical solution $\theta(t)$ of Eq.(\ref{eqexpmov}) linearized about $\theta=\pi$, with no fitting parameters except from $\xi=0.3$ and $\theta(0)=\pi+5 \times 10^{-3}$. In the limit where $\theta_n(t)\approx e^{-\xi t}\theta_{0\infty}(t)\Re(\rho^n)$ on each modulation period $n$, $\sigma_i^2[\theta(t)]$ is simply given by $|\theta_{i\infty}|^2$. The discrepancy between the experimental variance and the one given by $|\theta_{0\infty}(\tau)|^2$ of Eq.(\ref{eqmovBVP}) comes from the damping $\xi=0.3$ that is important in our system. When computing the numerical variance of the solution of Eq.(\ref{eqexpmov}) with $\xi=0$, the agreement becomes remarkable as shown with purple dots in Fig.\ref{fig4}A.

For  $B^+=933 \pm 5$ $\mu$T and $B^-=-1081 \pm 5$ $\mu$T, a second mode of stabilization is experimentally observed as displayed by the two trajectories of Fig.\ref{fig4}C. For these stable trajectories, that are again impossible to experimentally duplicate given the sensitivity to initial conditions, the compass is oscillating about $\theta=\pi$ with an angular response having two local maxima per period. The normalized variance of those trajectories is conserved and in good agreement with a numerical computation of Eq.(\ref{eqexpmov}) linearized about $\theta=\pi$ with only $\xi=0.3$ and $\theta(0)=\pi+5 \times 10^{-3}$ as fitting parameters (blue dots in Fig.\ref{fig4}A). Again, the discrepancy between the experimental variances and theoretical $|\theta_{1\infty}|^2$ is due to damping since a simulation with $\xi=0$ nicely match the second eigenfunction of Eq.(\ref{eqmovBVP}).

\section{Conclusions}

We have demonstrated a new minimal condition for the dynamic stability of systems with periodically time-varying parameters through the fundamental example of a mass in a local potential harmonic well whose curvature is varying between a positive and negative curvature. We have shown one can maintain stability while maximizing the time spend with a negative curvature as soon as the duration of positive curvature belongs to a discrete set of values that are remarkably predicted by the bound states of a quantum particle in a finite square potential well.

Optimal dynamical stabilization should offer a promising opportunity for a new type of passive control in dynamical systems whose design will be based on quantum-like rules such as quantization as shown in this article, but possibly superposition or electronic band gaps since those concepts have a physical interpretation in our framework. The proof of concept in this article has been done for a 1 degree-of-freedom system but it would be interesting to investigate what happens for more dynamical state variables. Also, we have focused on parameters varying in time but a similar approach should hold for parameters varying in space, allowing us for example to derive new minimal conditions for waves propagating in periodic media.

\begin{acknowledgments}
AL would like to thank the CNRS for the one year sabbatical award that allowed him the time to write this article. AL would also thanks S. Neukirch for orienting him towards Optimal control Theory and O. Devauchelle for invaluable discussions.
\end{acknowledgments}

\appendix

\section{Pontryagin Maximum Principle (PMP)}

Here we apply Pontryagin Maximum Principle to our linear mass-spring system where the stiffness is $T$-periodic in time. We recall the goal is to find the optimal control $u(t)$ that minimize the continuous-time functional $\int_0^Tu(t)dt$ with $u^- \leq u(t) \leq u^+$ subject to the first order dynamic constraints 
\begin{equation}
\left\{\begin{array}{ll}
\dot{x}(t)=y(t) \\
\dot{y}(t) = -u(t)x(t)
\end{array}\right.
\label{eqmovfirstorderannex}
\end{equation}
with $T>0$ and some boundary conditions to specify.

\textit{Theorem}: 
For the command $u(t)$ and the trajectory $\mathbf{x}(t)=\{x(t),y(t)\}^T$ to be a normal optimal, it is necessary that it exists a real $p^0 \leq 0$ and an adjoint vector function $\mathbf{p}(t)=\{p_x(t),p_y(t)\}^T$, so that the couple $(\mathbf{p}(t),p^0)$ should verify
\begin{equation}
\left\{\begin{array}{lll}
\dot{x}(t)= y(t) \\
\dot{y}(t)= -u(t)x(t) \\
\dot{p}_x(t)= p_y(t)u(t) \\
\dot{p}_y(t) = -p_x(t) \end{array}\right.
\label{Pontry1}
\end{equation}
with the maximum condition 
$$(p^0-p_y(t)x(t))u(t) = max_{(u^- \leq v \leq u^+)}\left[(p^0-p_y(t)x(t))v\right]$$
that is equivalent to 
\begin{equation}
u(t)=\left\{\begin{array}{ll}
u^+ \text{ if } p_y(t)x(t) < p^0\\
u^- \text{ if } p_y(t)x(t) > p^0 \\
\text{undefined if } p_y(t)x(t) = p^0  \end{array}\right.
\label{Pontry2}
\end{equation}
\\
We are interested in periodic boundary conditions so that $x(0)=x(T)$ and $y(0)=y(T)$. If $x(0)$ and $y(0)$ are free, the optimal solution of the PMP is the trivial one $x(t)=y(t)=0$ and $u(t)=u^-$ for all $t \in [0,T]$.  We are of course interested in non trivial solutions.

The problem in Eqs.(\ref{Pontry1})-(\ref{Pontry2}) being invariant by homothety (if $\left(x(t),y(t),u(t)\right)$ is an optimal solution then $\left(\mu x(t),\mu y(t),u(t)\right)$ with $\mu \neq 0$ is a solution), we can limit ourselves to the initial conditions $x(0)^2 + y(0)^2=1$. The problem then becomes to properly formulate the transversality condition of the PMP, a task that is not straightforward to achieve. A summary of the demonstration is shown here but the interested reader can find a complete detailed version of the theorem in \cite{Trelat2025}.

The first thing to prove when $x(0)^2 + y(0)^2=1$ is the periodicity of the adjoint $p_x(0)=p_x(T)$ and $p_y(0)=p_y(T)$ that  is proved in \cite{Trelat2025} but we will simply assume here. It is then possible to show \cite{Trelat2025} that there is an infinity of optimal solutions of Eqs.(\ref{Pontry1})-(\ref{Pontry2}) verifying $x(0)^2 + y(0)^2=1$ that are simply the same solution switched from $\alpha \in \mathbb{R}$ in time. One of those solutions corresponds to $x(0)=1$ and $y(0)=0$ (the associated optimal trajectory is unique) and can be considered the optimal solution without any loss of generality.  This solution has the following symmetry property $\forall t \in [0,T]$: $x(T-t)=x(t)$, $y(T-t)=-y(t)$, $u(T-t)=u(t)$, $p_x(T-t)=-p_x(t)$ and $p_y(T-t)=p_y(t)$. 

The next step in the demonstration comes from the assumptions on the control $u(t)$. If $u^-<0$ (and $p_y(0) \neq 0$), which is the case in this manuscript, we can prove \cite{Trelat2025} than $x(t)=-p_y(t)/|p_y(0)|$ and $y(t)=p_x(t)/|p_y(0)|$ $\forall t \in [0,T]$. Also, assuming $u^+>0$ and $\sqrt{u^+}T \not\in 2\pi \mathbb{N}^*$, we can demonstrate we are in the normal case, i.e. $p^0=-1$, and there is no singular arc, i.e. $u(t)$ is bang-bang. Thus, letting $\lambda=1/|p_y(0)|$, there is a unique optimal solution to Eqs.(\ref{Pontry1})-(\ref{Pontry2}), with $x(0)=x(T)=1$ and $y(0)=y(T)=0$, such that 
\begin{equation}
\left\{\begin{array}{ll}
\dot{x}(t)= y(t) \\
\dot{y}(t)= -u(t)x(t)\end{array}\right.
\label{Pontry1reducedannex}
\end{equation}
with a maximum condition 
\begin{equation}
u(t)=\left\{\begin{array}{ll}
u^+>0 \text{ if } x^2(t) > \lambda\\
u^-<0 \text{ if } x^2(t) < \lambda \end{array}\right.
\label{Pontry2reducedannex}
\end{equation}
where $\lambda>0$ is the only unknown of the reduced PMP problem.

Finally, we can also prove few more property of the optimal solution \cite{Trelat2025}. Notably, the control is bang-bang with two commutations
\begin{equation}
u(t)=\left\{\begin{array}{lll}
u^+ \quad \text{if }  \, 0 \leq t < T^+/2\\
u^- \quad \text{if } \, T^+/2 < t < T-T^+/2\\
u^+ \quad \text{if } \, T-T^+/2 < t < T \end{array}\right.
\label{Annexcontrol}
\end{equation}
Also, $x(t)$ is never zero. The optimal trajectory with $x(0)=x(T)=1$ and $y(0)=y(T)=0$ follows $0 < x_{min} \leq 1$ $\forall t \in [0,T]$ where $x_{min}$ depends on $T$ and converges towards $0$ when $T \rightarrow \infty$. Actually, in the large $T$ limit, it is straightforward to determine the duration $T^+$ of the bang-bang control $u(t)$ and the corresponding optimal solution $x(t)$, as shown in the following section.

\section{Optimal dynamical stabilization in the large $T$ limit}

\subsection{Optimal solution of the PMP}

When the period $T$ becomes large, the optimal $x(t)$ that minimize the quantity $\int_0^Tu(t)dt$ exhibits some turnpike properties: over long time $T$, ``cheapest'' trajectory is to approach the absolute minimum $x(t)=y(t)=0$ and $u(t)=u^-$ as long as possible. But to fulfill the boundary conditions $x(0)=x(T)=1$ and $y(0)=y(T)=0$, the trajectory has to be non-monotonic and the control $u(t)$ has to be $u^+>0$ during a time $T^+$. For $T \rightarrow \infty$ the solutions $x(t)$ and the unknown duration $T^+$, denoted $x_{\infty}(t)$ and $T_{\infty}^+$, respectively, are described by the ordinary differential equation
\begin{align}
\left\{\begin{array}{ll}
-\ddot{x}_{\infty}(\tau)=u^+x_{\infty}(\tau) & \text{for } | \tau | < T_{\infty}^+/2\\
-\ddot{x}_{\infty}(\tau)+(u^+-u^-)x_{\infty}(\tau)=u^+x_{\infty}(\tau) & \text{for } | \tau | > T_{\infty}^+/2 \end{array}\right.
\label{particleinpotentialwell}
\end{align}
with the boundary conditions $x_{\infty}(\tau) \rightarrow 0$ for $\tau\rightarrow -\infty$ and $\tau\rightarrow +\infty$. Since the PMP problem of Eq.(\ref{Pontry1reducedannex})-(\ref{Pontry2reducedannex})) is independent of the phase, we fixed the latter in Eq.(\ref{particleinpotentialwell}) by symmetrically centering the bang-bang control $u(\tau)$ about $\tau =0$. Doing so, the Boundary Value Problem in Eq.(\ref{particleinpotentialwell}) is directly analogous to the stationary Schrödinger equation of a particle in the $1D$ finite potential well $\tilde{u}(\tau)=0$ for $| \tau | < T_{\infty}^+/2$ and $\tilde{u}(\tau)=\Delta u = u^+-u^-$ for $| \tau | < T_{\infty}^+/2$ where space has been replaced with time. The following resolution is well-known in quantum physics. For a given width $T_{\infty}^+$ and height $\Delta u$ of the finite potential well, one is looking for an expression of the eigenvalues $u^+$ and eigenfunctions $x_{\infty}(\tau)$. In the PMP framework of this manuscript, $u^+$ and $u^- = \Delta u -u^+$ is fixed and we are looking for the optimal duration of the control $T_{\infty}^+$ and optimal response $x_{\infty}(\tau)$.

Since $u^+>0$ and $u^-<0$, the set of solutions $[u^+_i,x_{i\infty}]$ of Eq.(\ref{particleinpotentialwell}) is discrete and the eigenfunctions $x_{\infty}(\tau)$ have the general form
\begin{align}
\left\{\begin{array}{lll}
x^A_{\infty}(\tau) = Ge^{\sqrt{|u^-|} \tau} + Fe^{-\sqrt{|u^-|} \tau}  & \text{for } \tau < -T_{\infty}^+/2\\
x^B_{\infty}(\tau) = A\sin(\sqrt{u^+} \tau) + B\cos(\sqrt{u^+} \tau)  & \text{for } | \tau | < T_{\infty}^+/2\\
x^C_{\infty}(\tau) = He^{-\sqrt{|u^-|} \tau} + Ie^{\sqrt{|u^-|} \tau}  & \text{for } \tau > T_{\infty}^+/2\end{array}\right.
\label{infinitesolutions}
\end{align}
For the solutions $x_{\infty}(\tau)$ to be square integrable, we need $F=I=0$. The set of solutions approximating the PMP problem of Eq.(\ref{Pontry1reducedannex})-(\ref{Pontry2reducedannex})) when $T\rightarrow \infty$ is the so-called ground state in quantum mechanics that is the first symmetric eigenfunction $x_{0\infty}$ and eigenvalue $u^+_0$. Since the bound state is symmetric, it follows $A=0$ and $G=H$ in Eq.(\ref{infinitesolutions}). Finally, the continuity condition at $|\tau|=T_{\infty}^+/2$ for $x_{\infty}(\tau)$ and $\dot{x}_{\infty}(\tau)$ imposes
\begin{align}
\begin{array}{ll}
Ge^{-\sqrt{|u^-|}T^+_{\infty}/2} =  & B\cos(\sqrt{u^+}T^+_{\infty}/2) \\
-G\sqrt{u^-}e^{-\sqrt{|u^-|}T^+_{\infty}/2} =  & -\sqrt{u^+}B\sin(\sqrt{u^+}T^+_{\infty}/2) \end{array}
\label{continuitycondition}
\end{align}
Introducing the second line in the first line of Eq.(\ref{continuitycondition}), we get 
\begin{equation}
\label{energyequation}
\sqrt{|u^-|}=\sqrt{u^+}\tan(\sqrt{u^+}T^+_{\infty}/2)\end{equation}
This equation (called the energy equation in quantum mechanics) cannot be solved analytically but it allows to find the set of discrete $T^+_{i\infty}$ for a given $u^+$ and $u^-$, which is how our optimal problem is set up. The smallest of all the $T^+_{i\infty}$, denoted $T^+_{0\infty}$, is the one that approximate the PMP problem of Eq.(\ref{Pontry1reducedannex})-(\ref{Pontry2reducedannex})) in the large $T$ limit when $T\rightarrow \infty$.

The second information one get out of Eq.(\ref{continuitycondition}) is $B=Ge^{-\sqrt{u^-}T^+_{\infty}/2}/\cos(\sqrt{u^+}T^+_{\infty}/2)$ so that the symmetric eigenfunctions of Eq.(\ref{particleinpotentialwell}) reads
\begin{align}
\left\{\begin{array}{lll}
x^A_{i\infty}(\tau) = Ge^{\sqrt{|u^-|} \tau} & \text{for } \tau < -T_{i\infty}^+/2\\
x^B_{i\infty}(\tau) = G\frac{e^{-\sqrt{|u^-|}T^+_{i\infty}/2}}{\cos(\sqrt{u^+}T^+_{i\infty}/2)}\cos(\sqrt{u^+} \tau)  & \text{for } | \tau | < T_{i\infty}^+/2\\
x^C_{i\infty}(\tau) = Ge^{-\sqrt{|u^-|} \tau} & \text{for } \tau > T_{i\infty}^+/2\end{array}\right.
\label{boundstate}
\end{align}
The solution $x_{i\infty}(\tau)$ is defined up to one constant $G$, as expected since $x_{i\infty(\tau)}$ is the solution of Eq.(\ref{particleinpotentialwell}) that is a linear Ordinary Differential Equation. In particular, the boundary conditions we chose for the PMP problem in Eq.(\ref{Pontry1reducedannex})-(\ref{Pontry2reducedannex})) and illustrated in Fig.1 are $x(0)=x(T)=1$ and $y(0)=y(T)=0$ with the maximum condition $u(\tau)=u^+>0$ if $x^2(\tau)>\lambda$ and $u(\tau)=u^-<0$ if $x^2(\tau)<\lambda$. Introducing this information in Eq.(\ref{boundstate}) we get $G= \cos(\sqrt{u^+}T^+_{0\infty}/2) / e^{-\sqrt{|u^-|}T^+_{0\infty}/2}$ so the ground state reads 
\begin{align}
\left\{\begin{array}{lll}
x^A_{0\infty}(\tau) = \frac{\cos(\sqrt{u^+}T^+_{0\infty}/2)}{e^{-\sqrt{|u^-|}T^+_{0\infty}/2}}e^{\sqrt{|u^-|} \tau} & \text{for } \tau < -T_{i\infty}^+/2\\
x^B_{0\infty}(\tau) = \cos(\sqrt{u^+} \tau)  & \text{for } | \tau | < T_{i\infty}^+/2\\
x^C_{0\infty}(\tau) = \frac{\cos(\sqrt{u^+}T^+_{0\infty}/2)}{e^{-\sqrt{|u^-|}T^+_{0\infty}/2}}e^{-\sqrt{|u^-|} \tau} & \text{for } \tau > T_{i\infty}^+/2\end{array}\right.
\label{groundstate}
\end{align}
The solution $[T^+_{0\infty},x_{0\infty}(\tau)]$ given in Eq.(\ref{energyequation}) and Eq.(\ref{groundstate}) is the one shown throughout the manuscript where only the phase have been varied to match the purpose of each figure. The real $\lambda$ from the maximum condition one get at each commutation of the control at $\tau=T^+_{0\infty}/2$ shown in Fig.1B simply reads $\lambda_{\infty}=(\cos(\sqrt{u^+}T^+_{0\infty}/2))^2$ in the large $T$ limit, $T \rightarrow \infty$.

\subsection{Bound states of the stationary Schr\"odinger equation}

We focus here on the linear equation $\ddot{x}(t)+u(t)x(t)=0$, where $u(t)$ is the $T$-periodic modulation function with $u(t)=u^+>0$ during $T^+$ and $u(t)=u^-<0$ during the rest of the period $T^-=T-T^+$. In the large $T$ limit, the bounded solutions $x(t)$ and the unknown duration $T^+$, denoted $x_{\infty}(\tau)$ ans $T_{\infty}^+$, respectively, are described by the ordinary differential equation Eq.(\ref{particleinpotentialwell}), that can be recast in the more concise form 
\begin{equation}
\label{particleinpotentialwellconcise}
\left(-\frac{d^2}{d\tau^2}+ \tilde{u}(\tau)\right)x_{\infty}(\tau)=u^+x_{\infty}(\tau)
\end{equation}
where
\begin{align}
\left\{\begin{array}{lll}
\tilde{u}(\tau) = u^+-u^- = \Delta u & \text{for } \tau < -T_{\infty}^+/2\\
\tilde{u}(\tau) = 0  & \text{for } |\tau| < T_{\infty}^+/2\\
\tilde{u}(\tau) = u^+-u^- = \Delta u  & \text{for } \tau > T_{\infty}^+/2\end{array}\right.
\label{finitepotentialwell}
\end{align}
where we recall the local time $\tau$ starts at the middle of each $T^+_{\infty}$ region and $x_{\infty}(\tau)$ is compact on each period, i.e. $x(\tau) \rightarrow 0$ for $\tau\rightarrow -\infty$ and $\tau\rightarrow +\infty$. The Boundary Value Problem in Eq.(\ref{particleinpotentialwellconcise}) is mathematically analogous to the stationary Schrödinger equation of a particle in a $1D$ finite potential well and its resolution is well-known. 

Since $u^+>0$ and $u^-<0$, the solutions of Eq.(\ref{particleinpotentialwellconcise}) have the general form
\begin{align}
\left\{\begin{array}{lll}
x^A_{\infty}(\tau) = Ge^{\sqrt{|u^-|} \tau} + Fe^{-\sqrt{|u^-|} \tau}  & \text{for } \tau < -T_{\infty}^+/2\\
x^B_{\infty}(\tau) = A\sin(\sqrt{u^+} \tau) + B\cos(\sqrt{u^+} \tau)  & \text{for } | \tau | < T_{\infty}^+/2\\
x^C_{\infty}(\tau) = He^{-\sqrt{|u^-|} \tau} + Ie^{\sqrt{|u^-|} \tau}  & \text{for } \tau > T_{\infty}^+/2\end{array}\right.
\label{infinitesolutions2}
\end{align}
The set of solutions $[u^+,x_{\infty}]$ of Eq.(\ref{particleinpotentialwellconcise}) is discrete when $u^+>0$ and $u^-<0$ as it is the case in this manuscript and we denote $[u^+_i,x_{i\infty}(t)]$ the $i^{th}$ set of solutions. The set of solutions approximating the PMP problem of Eq.(\ref{Pontry1reducedannex})-(\ref{Pontry2reducedannex})) when $T\rightarrow \infty$ are the width of the finite potential well $T^+_{0\infty}$ and the eigenfunction $x_{0\infty}(t)$ associated with the smallest $u^+$, called the ground state in quantum mechanics.

For the solutions $x_{\infty}(\tau)$ to be square integrable, we need $F=I=0$. Next, we know that the $x_{\infty}(\tau)$ function must be continuous and differentiable.  In other words, the values of the functions and their derivatives must match up at the dividing points:
\begin{equation} 
\left \{
\begin{split}
&x^A_{\infty}(-T_{\infty}^+/2)=x^B_{\infty}(-T_{\infty}^+/2) \text{,} \quad x^B_{\infty}(T_{\infty}^+/2)=x^C_{\infty}(T_{\infty}^+/2)  \\
&\dot{x}^A_{\infty}(-T_{\infty}^+/2)=\dot{x}^B_{\infty}(-T_{\infty}^+/2) \text{,} \quad \dot{x}^B_{\infty}(T_{\infty}^+/2)=\dot{x}^C_{\infty}(T_{\infty}^+/2)  
\end{split}
\right.
\label{continuity2}
\end{equation}
These equations have two sort of solutions, symmetric, for which $A=0$ and $G=H$ (like the optimal solution or ``ground state'' associated to $T^+_{0\infty}$) and antisymmetric, for which $B=0$ and $G=-H$.

The symmetric solutions have already be given in Eqs.(\ref{continuitycondition})-(\ref{boundstate}) to get a semi-analytical and analytical expression of $T^+_{i\infty}$ and $x_{i\infty}$, respectively. The continuity conditions Eq.(\ref{continuity2}) give, for antisymmetric solutions
\begin{align}
\begin{array}{ll}
Ge^{-\sqrt{|u^-|}T^+_{\infty}/2} =  & -A\sin(\sqrt{u^+}T^+_{\infty}/2) \\
G\sqrt{u^-}e^{-\sqrt{|u^-|}T^+_{\infty}/2} =  & \sqrt{u^+}A\cos(\sqrt{u^+}T^+_{\infty}/2) \end{array}
\label{continuitycondition2}
\end{align}
Introducing the second line in the first line of Eq.(\ref{continuitycondition2}), we get 
\begin{equation}
\label{energyequation2}
\sqrt{|u^-|}=-\sqrt{u^+}/\tan(\sqrt{u^+}T^+_{\infty}/2)\end{equation}
This equation cannot be solved analytically but it allows to find the set of discrete $T^+_{i\infty}$ associated with antisymmetric solutions $x_{i\infty}(\tau)$ such as the one given in Fig.2E. 

The second information one get out of Eq.(\ref{continuitycondition2}) is $A=-Ge^{-\sqrt{u^-}T^+_{\infty}/2}/\sin(\sqrt{u^+}T^+_{\infty}/2)$ so that the antisymmetric bound states reads
\begin{align}
\left\{\begin{array}{lll}
x^A_{i\infty}(\tau) = Ge^{\sqrt{|u^-|} \tau} & \text{for } \tau < -T_{i\infty}^+/2\\
x^B_{i\infty}(\tau) = -G\frac{e^{-\sqrt{|u^-|}T^+_{i\infty}/2}}{\sin(\sqrt{u^+}T^+_{i\infty}/2)}\sin(\sqrt{u^+} \tau)  & \text{for } | \tau | < T_{i\infty}^+/2\\
x^C_{i\infty}(\tau) = -Ge^{-\sqrt{|u^-|} \tau} & \text{for } \tau > T_{i\infty}^+/2\end{array}\right.
\label{boundstate2}
\end{align}
The solution $x_{i\infty}(\tau)$ is defined up to one constant $G$ The solution $[T^+_{1\infty},x_{1\infty}(\tau)]$ given in Eq.(\ref{energyequation2})-(\ref{boundstate2}) is the one shown in Fig.1E and Fig.4A.

\begin{figure}[!h]
\centering
\includegraphics[width=0.83\columnwidth]{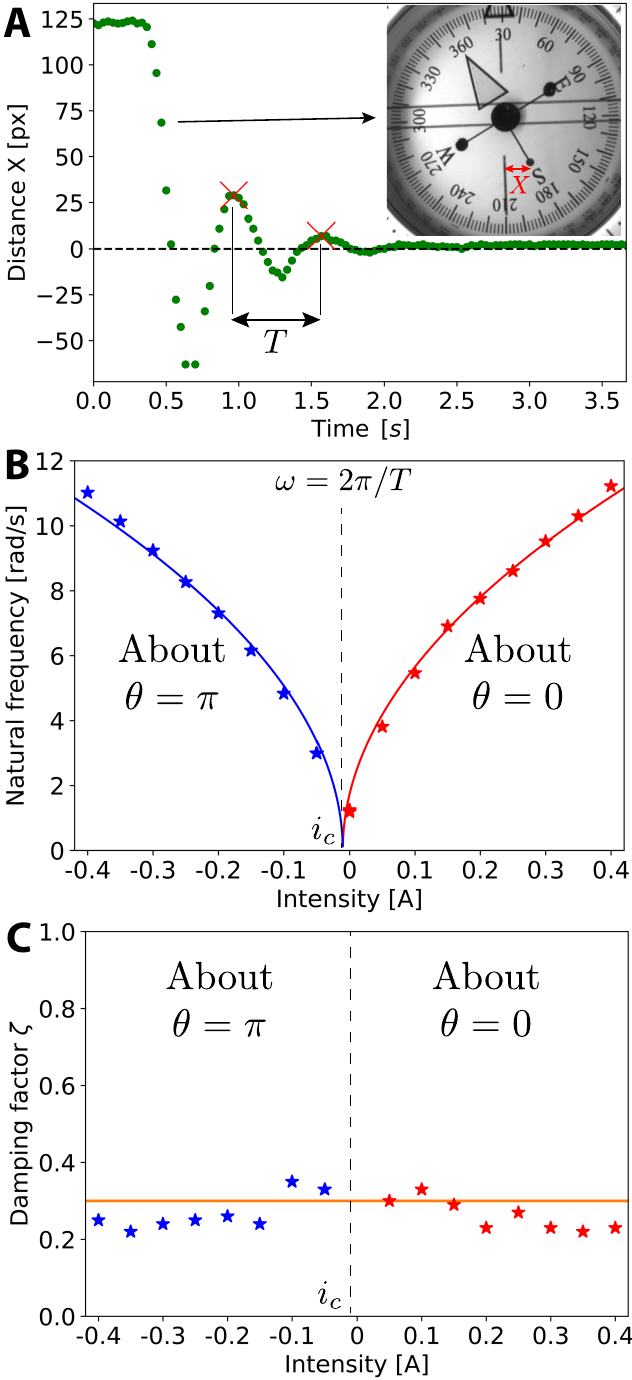}
\caption{Experimental characterization of the linear oscillatory behavior of the compass as a function of current $i$ in the Helmoltz coils. A. Example of  damped oscillations of the south pole of the compass about $\theta = \pi$ for $i=-0.35$ A. The ``pseudo-natural period'' $T$ is obtained by measuring the duration between two subsequent crest in the compass's oscillations. Inset shows a top picture of the compass at a given time. B. Evolution of the pseudo natural frequency $\omega = 2\pi/T$ of the free vibrating compass as a function of current $i$. Above a critical current $i=i_c=-0.01$ A, the south pole of the compass oscillates about $\theta = 0$ (geographic south). Below $i=i_c$, the stability is upside-down and the south pole of the compass oscillates about $\theta = \pi$. The solid lines are the theoretical predictions leading to a measured $\mu/I=6.4 \times 10^4$ A.kg$^{-1}$. C. Evolution of the damping ratio $\xi$ as a function of $i$. The damping ratio is measured through the classic logarithmic decrement method applied on the two successive peaks of Fig. A.}
\label{figA}
\end{figure}

\section{Experimental characterization of the compass}

For the governing equation of the motion of the compass inside the two Helmhotlz to be complete, we measure the damping ratio $\xi$ and the ratio between the magnetic moment $\mu$ and moment of inertia $I$ through the classic characterization experiments depicted in Figure \ref{figA}. The experiment consists in carefully orienting a compass along the main axis of two Helmholtz coils that are themselves collinear to the earth magnetic field. Doing so, the $S-N$ axis of the compass is collinear to a uniform unidirectional magnetic field that is the sum of the earth magnetic field and a magnetic field eventually generated by a current $i$ in the coils (this setup is the one used to simply measure the earth magnetic field). When $i=0$, the south pole of the compass points toward the south geographic pole that is up when looking at the picture of the inset of Figure \ref{figA}A (this equilibrium position is denoted $\theta = 0$). When $i>i_c=-0.01$ A, the total magnetic field becomes upside-down and the south pole of the compass points down towards the geographic north, the stable equilibrium is now $\theta = \pi$.

The initial condition of the characterization experiments consists, for a given current $i$ in the coils, in bringing the south pole of the compass away from its stable position thanks to an electromagnet. We then turn off the electromagnet and we record the motion of the compass about its stable equilibrium position. Monitoring the distance $X$ between the south pole of the compass and the geographic South-North axis, we can mesure the pseudo natural frequency and the damping ratio of the damped oscillatory motion. An example of an experimental oscillatory motion about the equilibrium position $\theta = \pi$ is shown in Figure \ref{figA}A for $i=-0.35$ A. For each $i$, one can extract a pseudo natural period $T$, and therefore a pseudo natural frequency $\omega=2\pi/T$ as shown in Figure \ref{figA}B (the red stars correspond to oscillations about $\theta =0$ when $i> i_c$ and the blue stars correspond to oscillations about $\theta =\pi$ when $i< i_c$.). Taking $B_T=47$ $\mu$T for the earth magnetic field in our lab and $A=(\frac{4}{5})^{(3/2)}\mu_0N/r=4496$ $\mu$T/A a property of our Helmoltz coils with $\mu_0=4\pi\times 10^{-7}$ $Tm/A$ the permeability of vacuum, $N=500$ the number of coils and $r=10$ cm the radius of the coils, those experimental pseudo natural frequencies $\omega$ best follow the theoretical curve
\begin{equation}
\label{eqannexcarac}
\omega(i)=\sqrt{(B_T+Ai)\frac{\mu}{I}}\end{equation}
when the ratio $\mu/I$ is chosen to be $\mu/I=6.4\times 10^4$ A.kg$^{-1}$. The theoretical curves given in Eq.(\ref{eqannexcarac}) are plotted with continuous blue and red curves about $\theta=\pi$ and $\theta=0$, respectively. 

A measure of the damping ration $\xi$ of the compass is shown in Figure \ref{figA}C. By fitting a constant to measure of the damping ratios for various $i$, we find that $\xi=0.3$ is a fair approximation to take in our numerical model of Eq.(\ref{eqexpmov}). The damping ratios have been measured on experimental data such as the one in Figure \ref{figA}A using the decrement logarithmic method
\begin{equation}
\label{eqannexloga}
\xi=\frac{\delta}{\sqrt{4\pi^2+\delta^2}}\end{equation}
where the logarithmic decrement reads $\delta = \ln(X(t)/X(t+T))$ and where $X(t)$ and $X(t+T)$ are the amplitudes of two successive peaks illustrated by red crosses in Figure \ref{figA}A. The logarithmic decrement method is usually adequate for lightly damped systems with $\xi < 0.1$ but we will see in this manuscript that the value of $\xi=0.3$ is doing a good job in predicting the stable responses and associated modulation functions of our experimental system.

\begin{figure*}[!t]
\centering
\includegraphics[width=0.95\textwidth]{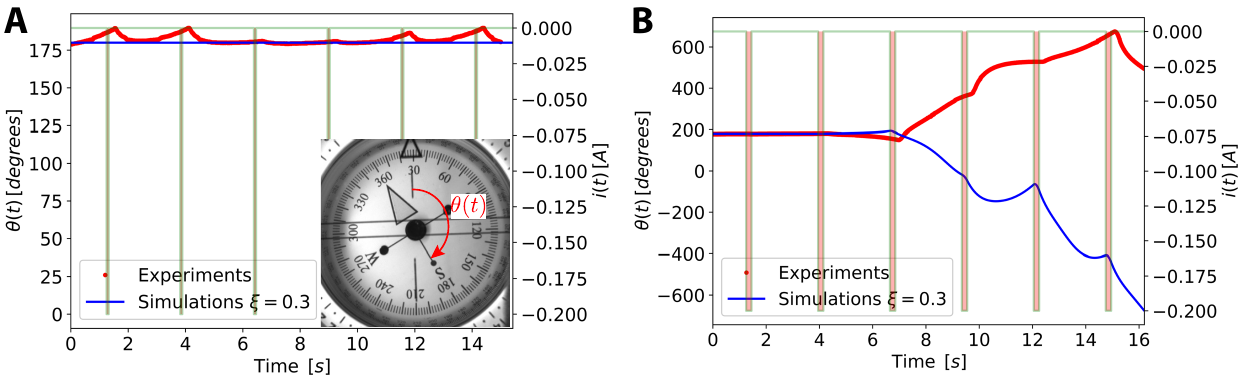}
\caption{Experimental and numerical results showing the angular position $\theta(t)$ of the south pole of the compass ($\theta(t)=\pi$ is the south pole of the compass pointing towards the magnetic south pole of earth that is naturally unstable if $i(t)=0$, i.e. there is no current in the Helmhotlz coils) as a function of time for $6$ periods of current modulation $i(t)$ and initial conditions $\theta(0)=\pi$ and $\dot{\theta}(0)=0$. In this experiment, we turn off the current in the Helmhotlz coils ($i(t)=0$) during $T^-=2.5$ s and we put a current $i(t)=-0.2$ A during a duration $T^+$ that we vary across different experiments. The period of the modulated current is $T=T^-+T^+$ and the modulated current $i(t)=i(t+T)$ is shown with light green lines (the quantity $\int_0^{6T}i(t)dt$ is highlighted in light red). The red dots are experimental $\theta(t)$ and the blue line is the one obtained by solving Eq.(9) of the manuscript with a damping ratio $\xi=0.3$. In the numerics, the initial conditions are $\theta(0)=\pi + 5\times 10^{-3}$ (a fitting parameter coming from figure 3B of the manuscript) and $\dot{\theta}(0)=0$ A. $T^+=70$ ms. B. $T^+=200$ ms.}
\label{figB}
\end{figure*}

\begin{figure*}[!t]
\centering
\includegraphics[width=0.95\textwidth]{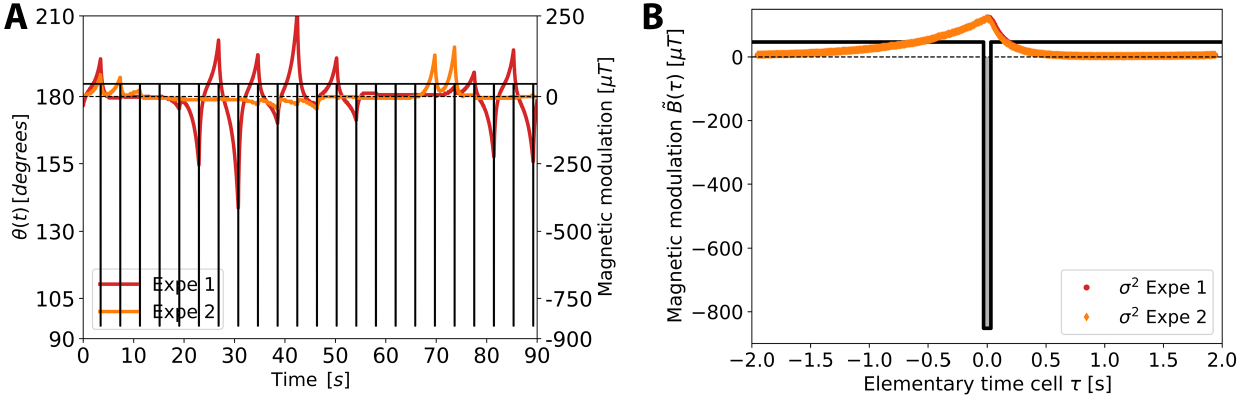}
\caption{Two experiments showing the angular position of the south pole of the compass $\theta(t)$ during $90$ seconds for $B(t) =\mathbf{B}(i)\mathbf{e}_x=B^-=-47$ $\mu$T ($i(t)$=0 in the coils) during $T^-=3.8$ s and $B(t) =B^+=+852$ $\mu$T during $T^+=70$ ms. A. Experimental $\theta(t)$ of the two experiments along with the amplitude of the modulated magnetic field $B(t)=B(t+T)$ where $T=T^-+T^+$. Although the two experiments have been realized for the same initial conditions $\theta(0)=\pi$ and $\dot{\theta}(0)=0$ and the same modulated parameters, the experimental outcome is different because the compass is undergoing a symmetry breaking every period of modulation, which makes it sensitive to initial conditions. B. Here we plot the variance $\sigma^2(t)$ of the angular position $\theta(t)$ within the modulation period (the period is centered about the pulse of $B^+$) in red for the first experiment and yellow for the second one.}
\label{figC}
\end{figure*}

\section{Experiments in the optimal stability region}

In this section, we give supplementary informations regarding figures 3B and 3C of the manuscript about the stabilization of the south pole of the compass towards the magnetic south pole of the earth while minimizing the quantity $\int i(t)dt$ where $i(t)$ is the current in the Helmholtz coils. In the first experiment described in figures S2 and 3B of the manuscript, we fix $i(t)=0$ ($B(t) =\mathbf{B}(i)\mathbf{e}_x=B^-=-47$ $\mu$T) during $T^-=2.5$ s and $i(t)=-0.2$ A ($B(t) =B^+=+852$ $\mu$) during $T^+$. For initial conditions $\theta(0)=\pi$ and $\dot{\theta}(0)=0$ and for various $T^+$, we look at the angular response $\theta(t)$  during 6 periods of modulation $T=T^-+T^+$. In figure S2 A for $T^+=70$ ms, the compass is dynamically stabilized and the south pole of the needle slightly oscillates about $\theta=\pi$ ($\theta$ is represented in the inset of the figure for information) as represented by red dots. This experimental results is in good qualitative agreement with the numerical data in blue line that ave been obtained by solving 
\begin{equation}
\label{eqexpmov}
\ddot{\theta}(t)+2\xi \sqrt{\frac{|\mathbf{B}(i)\mathbf{e}_x|\mu}{I}}\dot{\theta}(t)+\frac{\mathbf{B}(i)\mathbf{e}_x\mu}{I}\sin(\theta(t))=0\end{equation}
with $\xi=0.3$, $\mu/I=6.4 \times 10^4$ A.kg$^{-1}$ and $\mathbf{B}(i)=-(B_T + A i)\mathbf{e}_x$ with $B_T=47$ $\mu$T and $A=4496 $ $\mu$T/A with initial conditions $\theta(0)=\pi+5\times 10^{-3}$ and $\dot{\theta}(0)=0$. The experimental and numerical results both agree that $T^+=70$ ms corresponds to a dynamic stabilization of the compass in its upside-down equilibrium state and lead to similar quantity  $\Delta \mathcal{W} = \int_0^{6T}|\theta(t) B(t)sin(\theta(t))|dt$ as shown in figure 3B of the manuscript but a more quantitative agreement is not possible because the numerical $\theta(t)$ will eventually converges to $\theta(t)=\pi$ for large $t$ when the experimental one is always oscillating about $\theta=\pi$ due to symmetry breaking when $\theta(t) \approx \pi$ and $\dot{\theta}(t) \approx 0$ and $\theta(t) = \pi$ is unstable for $i(t)=0$. Figure S2 B shows the same experiment but for $T^+=200$ ms. This time, both the experimental and numerical angular response $\theta(t)$ dynamically diverges from $\theta(t)=\pi$ confirming that $\theta(t)=\pi$ is unstable for this $T^+$. Again, the numerics and experiments qualitatively agree, notably on the quantity of work done by the compass in the magnetic field represented by $\Delta \mathcal{W} = \int_0^{6T}|\theta(t) B(t)sin(\theta(t))|dt$, but quantitatively  disagree on the angular position $\theta(t)$, even after few modulation periods, because again of the symmetry breaking nature of the motion.

We focus now on the particular case where $T^+=70$ ms and try to experimentally increase $T^-$ until the south pole of the compass is not dynamically stabilized on the magnetic south pole of the earth anymore. The highest $T^-$ we obtained was $T^-=3.8$ s ($i(t) = 0$ $98 \%$ of the time)  and the corresponding angular position $\theta(t)$ about $\theta = \pi$, recorded during $90$ s, is given in figure S3. Figure S3 A, shows the evolution of $\theta(t)$, along with the amplitude of the magnetic field $B(t)$ for two experiments realized with the same conditions and $B(t)$. Again, because of the periodic symmetry breaking nature of the motion, we cannot reproduce the same experiments and the two experimental $\theta(t)$ of figure S3 A looks different, almost already from the first modulation period. What is conserved throughout experiments though, is the shape of the oscillations about $\theta=\pi$ on each modulation period, which is an inherent property of optimal dynamical stabilization we have described in the theoretical part of the manuscript. This property is observed when one plots the variance $\sigma^2(t)$ of $\theta(t)$ on each period as shown in figure S3 B where we chose a pulse-centered elementary periodic cell. For the optimal mode about $T^+_{0\infty}$, the one corresponding to the ground state of the quantum analog, the dispersion of the oscillations $\theta(t)$ is higher at the middle of the pulse and almost zero far away in time form it because then, the south ole of the needle is almost always close to $\pi$. We will show in figure 4 of the manuscript that the square of the absolute value of the eigenfunctions $\theta_{n\infty(\tau)}$ of the stationary Schrödinger equation (8) of the manuscript gives a good prediction of the variance of the dynamically stabilized motions.

\FloatBarrier
\vspace{1cm}

Movie S1: Movie associated to figure S1.A. It shows the top view of the damped oscillations of the south magnetic pole of the compass about $\theta = \pi$ for $i=-0.35$ A when released from an initial conditions.

Movie S2: Movie associated to figure S2.A, showing the angular position $\theta(t)$ of the south pole of the compass as a function of time for $6$ periods of current modulation $i(t)$ and initial conditions $\theta(0)=\pi$ and $\dot{\theta}(0)=0$. In this experiment, we turn off the current in the Helmhotlz coils ($i(t)=0$) during $T^-=2.5$ s and we put a current $i(t)=-0.2$ A during a duration $T^+=70$ ms (corresponding to the flashes of light).

Movie S3: Same as movie S1.A but associated to figure S2.B with $T^+=200$ ms.

Movie S4: Movie associated with the first experiment (labeled espe1) of figure S3 showing the time evolution of the angular position of the south pole of the compass $\theta(t)$ during $90$ seconds for $B(t) =\mathbf{B}(i)\mathbf{e}_x=B^-=-47$ $\mu$T ($i(t)$=0 in the coils) during $T^-=3.8$ s and $B(t) =B^+=+852$ $\mu$T during $T^+=70$ ms. The flashing lights corresponds to the durations $T^+$.

Movie S5: Same as movie S4 but we repeat the experiment a second time (labeled expe 2 in figure S3).

Movie S6: Movie associated to the first experiment (labeled expe 1) of figure 4B of the manuscript showing the time evolution of the angular position of the first mode of the south magnetic pole of the compass $\theta(t)$ during $50$ seconds for $T^+=630$ ms, $T^-=320$ ms, $B^+=245 \pm 5$ $\mu$T and $B^-=-1746 \pm 5$  $\mu$T. Right light is the duration $T^+$ when left light is $T^-$.

Movie S7: Same as movie S6 but we repeat the experiment a second time (labeled expe 2 in figure 4B of the manuscript). Note this movie duration was only $11$ seconds because of camera saving problems.

Movie S8: Movie associated to the first experiment (labeled expe 1) of figure 4C of the manuscript showing the time evolution of the angular position of the second mode of the south magnetic pole of the compass $\theta(t)$ during $50$ seconds for $T^+=630$ ms, $T^-=320$ ms, $B^+=933 \pm 5$ $\mu$T and $B^-=-1081 \pm 5$ $\mu$T. Right light is the duration $T^+$ when left light is $T^-$.

Movie S9: Same as movie S8 but we repeat the experiment a second time (labeled expe 2 in figure 4C of the manuscript).

\nocite{*}

\bibliography{PRA_Researcharticle_Lazarus_Trelat}

\end{document}